\documentclass[11pt]{article}
\usepackage{graphics}
\newcommand{\be}{\begin{equation}}
\newcommand{\ee}{\end{equation}}
\newcommand{\s}{|s\rangle}
\newcommand{\lx}{\langle x|}
\newcommand{\rx}{|x\rangle}
\newcommand{\x}{|x\rangle}

\newcommand{\ls}{\langle s|}
\newcommand{\rs}{|s\rangle}
\newcommand{\lu}{\langle u|}
\newcommand{\ru}{|u\rangle}
\newcommand{\lut}{\langle u(t)|}
\newcommand{\rut}{|u(t)\rangle}
\newcommand{\binent}{{\cal{H}}}

\newtheorem{Proposition}{Proposition}
\newtheorem{Lemma}{Lemma}

\begin{document}
\thispagestyle{empty}
\title{\bf An information-theoretic analysis of Grover's algorithm}
\author{{Erdal Arikan}\\
Electrical-Electronics Engineering Department,\\
Bilkent University, 06533 Ankara, Turkey.\\
{\rm arikan@ee.bilkent.edu.tr}}
\date{\today}
\maketitle
\begin{abstract}
Grover discovered a quantum algorithm for identifying a target element in an unstructured search universe of $N$ items in approximately $\pi/4 \sqrt{N}$ queries to a quantum oracle. 
For classical search using a classical oracle, the search complexity is
of order $N/2$ queries since on average half of the items must be searched.
In work preceding Grover's, Bennett et al. had shown that no quantum algorithm can solve the search problem in fewer than $O(\sqrt{N})$ queries.
Thus, Grover's algorithm has optimal order of complexity. 
Here, we present an information-theoretic analysis of Grover's 
algorithm and show that the square-root speed-up by Grover's algorithm is the best possible by any algorithm using the same quantum oracle.
\end{abstract}


\section{Introduction}

Grover \cite{Grover}, \cite{Grover2} discovered a quantum algorithm for identifying a target element in an unstructured search universe of $N$ items in approximately $\pi/4 \sqrt{N}$ queries to a quantum oracle. 
For classical search using a classical oracle, the search complexity is clearly
of order $N/2$ queries since on average half of the items must be searched.
It has been proven that this square-root speed-up is the best attainable
performance gain by any quantum algorithm.
In work preceding Grover's, Bennett et al. \cite{Bennett} had shown that no quantum algorithm can solve the search problem in fewer than $O(\sqrt{N})$ queries.
Following Grover's work, Boyer et al. \cite{Boyer} showed that Grover's algorithm is optimal asymptotically, and that square-root speed-up cannot be improved even if one allows, e.g., a 50\% probability of error. Zalka \cite{Zalka} strengthened these results to show that Grover's algorithm is optimal exactly (not only asymptotically).
In this correspondence we present an information-theoretic analysis of Grover's 
algorithm and show the optimality of Grover's algorithm from a different point of view.

\section{A general framework for quantum search \label{generalframework}}
We consider the following general framework for quantum search algorithms.
We let $X$ denote the state of the target and $Y$ the output of the search algorithm.
We assume that $X$ is uniformly distributed over the integers $0$ through $N-1$.
$Y$ is also a random variable distributed over the same set of integers.
The event $Y=X$ signifies that the algorithm correctly identifies the target.
The probability of error for the algorithm is defined as $P_e=P(Y\neq X)$.

The state of the target is given by the density matrix
\be
\rho_T = \sum_{x=0}^{N-1} (1/N) \rx \lx,
\label{targetstate}
\ee
where $\{\rx\}$ is an orthonormal set.
We assume that this state is accessible to the search algorithm only through calls to an oracle whose exact specification will be given later.
The algorithm output $Y$ is obtained by a measurement performed on the state of the quantum computer at the end of the algorithm.
We shall denote the state of the computer at time $k=0,1,\ldots$ by the density matrix $\rho_C(k)$. 
We assume that the computation begins at time 0 with the state of the computer given by an initial state $\rho_C(0)$ independent of the target state.
The computer state evolves to a state of the form 
\be
\rho_C(k) = \sum_{x=0}^{N-1} (1/N)\rho_x(k)
\label{computerstate}
\ee
at time $k$, under the control of the algorithm.
Here, $\rho_x(k)$ is the state of the computer at time $k$, conditional on the target value being $x$. 
The joint state of the target and the computer at time $k$ is given by 
\be
\rho_{TC}(k) = \sum_{x=0}^{N-1} (1/N) \rx \lx \otimes \rho_x(k).
\label{jointstate}
\ee
The target state (\ref{targetstate}) and the computer state (\ref{computerstate}) can be obtained as partial traces of this joint state.

We assume that the search algorithm consists of the application of a sequence of unitary operators on the joint state. Each operator takes one time unit to complete.
The computation starts at time 0 and terminates at a predetermined time $K$, when a measurement is taken on $\rho_C(K)$ and $Y$ is obtained.
In accordance with these assumptions, we shall assume that the time index $k$ is an integer in the range 0 to $K$, unless otherwise specified.

There are two types of unitary operators that may be applied to the joint state by a search algorithm: oracle and non-oracle.
A non-oracle operator is of the form $I\otimes U$ and acts on the joint state as
\be
\rho_{TC}(k+1) =  (I \otimes U)\, \rho_{TC}(k)\, (I\otimes U)^\dagger =
\sum_{x} (1/N) |x\rangle \langle x| \otimes U \rho_x(k) U^\dagger.
\label{nonoraclejoint}
\ee
Under such an operation the computer state is transformed as
\be
\rho_C(k+1) = U \rho_C(k)U^\dagger.
\label{nonoraclecomputer}
\ee
Thus, non-oracle operators act on the conditional states $\rho_x(k)$ uniformly; $\rho_x(k+1) = U\rho_x(k)U^\dagger$.
Only oracle operators have the capability of acting on conditional states non-uniformly.

An oracle operator is of the form $\sum_x |x\rangle\langle x| \otimes O_x$ and 
takes the joint state $\rho_{TC}(k)$ to 
\be
\rho_{TC}(k+1) =  
\sum_x (1/N) \rx \lx \otimes O_x \rho_x(k) O_x^\dagger.
\ee
The action on the computer state is
\be
\rho_C(k+1) = \sum_x (1/N) O_x \rho_x(k) O_x^\dagger.
\label{oraclecomputer}
\ee

All operators, involving an oracle or not, preserve the entropy of the joint state $\rho_{TC}(k)$.
The von Neumann entropy of the joint state remains fixed at
$S[\rho_{TC}(k)]=\log N$ throughout the algorithm.
Non-oracle operators preserve also the entropy of the computer state;
the action (\ref{nonoraclecomputer}) is reversible, hence 
$S[\rho_C(k+1)] = S[\rho_C(k)]$.
Oracle action on the computer state (\ref{oraclecomputer}), however, does not preserve entropy; $S[\rho_C(k+1)]\neq S[\rho_C(k)]$, in general.

Progress towards identifying the target is made only by oracle calls that have the capability of transferring information from the target state to the computer state. We illustrate this information transfer in the next section.

\section{Grover's algorithm}
Grover's algorithm can be described within the above framework as follows.
The initial state of the quantum computer is set to 
\be
\rho_C(0) = \rs \ls
\label{groverinitialstate}
\ee
where
\be
\rs = \sum_{x=0}^{N-1} (1/\sqrt{N})\rx.
\label{meanstate}
\ee
Since the initial state is pure, the conditional states $\rho_x(k)$ will also be pure
for all $k\ge 1$.

Grover's algorithm uses two operators: an oracle operator with 
\be
O_x = I - 2 \rx \lx,
\label{oracleoperator}
\ee
and a non-oracle operator (called `inversion about the mean')   
given by $I\otimes U_s$ where
\be
U_s = 2 \rs \ls - I.
\ee
Both operators are Hermitian.

Grover's algorithm interlaces oracle calls with inversion-about-the-mean operations.
So, it is convenient to combine these two operations in a single operation, called Grover iteration,
by defining $G_x=U_s O_x$. 
The Grover iteration takes the joint state $\rho_{TC}(k)$ to
\be
\rho_{TC}(k+1) = \sum_x (1/N) \rx \lx \otimes G_x \rho_x(k) G_x^\dagger
\label{groveriteration}
\ee
In writing this, we assumed, for notational simplicity, that $G_x$ takes one time unit to complete, although it consists of the succession of two unit-time operators.

Grover's algorithm consists of $K=(\pi/4) \sqrt{N}$ successive applications of Grover's iteration 
beginning with the initial state (\ref{groverinitialstate}), followed by a  measurement on $\rho_C(K)$ to obtain $Y$.
The algorithm works because the operator $G_x$ can be interpreted as a rotation of the $x$--$s$ plane by an angle $\theta = \arccos(1-2/N)\approx 2/\sqrt{N}$ radians.
So, in $K$ iterations, the initial vector $\s$, which is almost orthogonal to $\x$, is brought into alignment with $\x$.

\begin{figure}[htb]
\begin{center}
\resizebox{!}{4in}{
\includegraphics{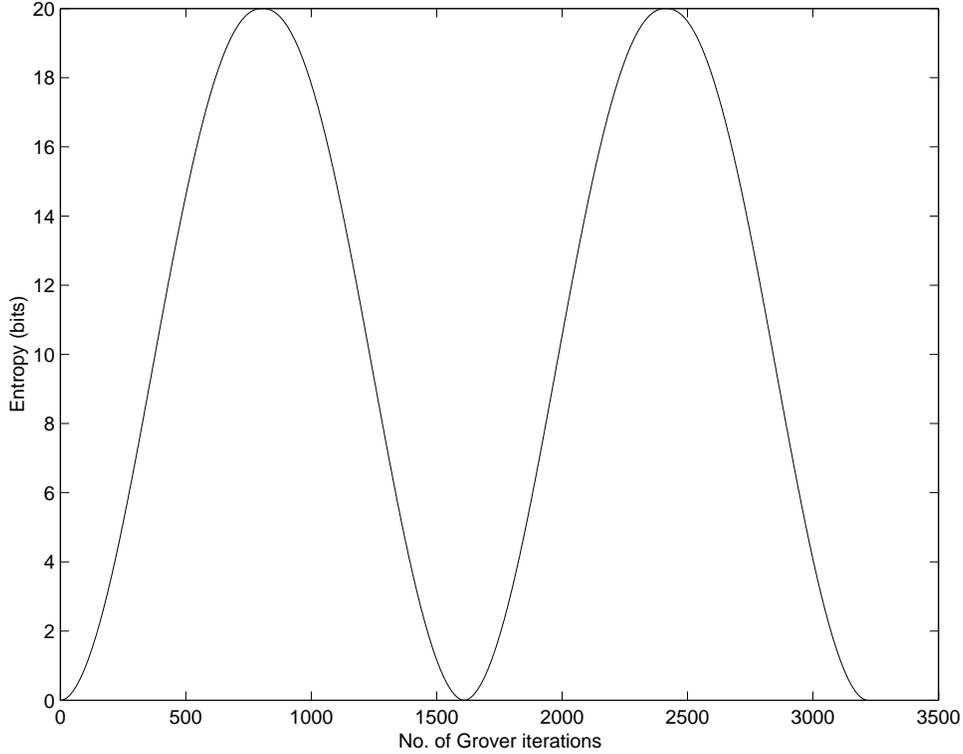}
}
\caption{Evolution of entropy in Grover's algorithm.}
\label{entropy.fig}
\end{center}
\end{figure}

Grover's algorithm lends itself to exact calculation of the eigenvalues of 
$\rho_C(k)$, hence to computation of its entropy.
The eigenvalues of $\rho_C(k)$ are
\be
\lambda_1(k)= \cos^2(\theta k)
\ee
of multiplicity 1, and
\be
\lambda_2(k)=\frac{\sin^2(\theta k)}{N-1}
\ee
of multiplicity $N-1$.
The entropy of $\rho_C(k)$ is given by
\be
S(\rho_C(k)) = -\lambda_1(k) \log \lambda_1(k)  - (N-1)\lambda_2(k) \log \lambda_2(k)
\ee
and is plotted in Fig.~\ref{entropy.fig} for $N=2^{20}$.
(Throughout the paper, the unit of entropy is bits and $\log$ denotes base 2 logarithm.)
The entropy $S(\rho_c(k))$ has period $\pi/\theta \approx (\pi/2)\sqrt{N}$. 

Our main result is the following lower bound on time-complexity.
\begin{Proposition}
Any quantum search algorithm that uses the oracle calls $\{O_x\}$ as defined by
(\ref{oracleoperator}) must call the oracle at least
\be
K \ge \left(\frac{1- P_e}{2\pi} +  \frac{1}{\pi \log N}\right) \sqrt{N} 
\label{Prop1}
\ee
times to achieve a probability of error $P_e$.
\end{Proposition}

For the proof we first derive an information-theoretic inequality.
For any quantum search algorithm of the type described in section~\ref{generalframework}, we have by Fano's inequality, 
\be
H(Y|X) \le \binent(P_e) + P_e \log(N-1) \le \binent(P_e) + P_e \log(N),
\label{Fano}
\ee
where for any $0\le u\le1$
\be
\binent(u) = - \delta \log \delta -(1-\delta) \log(1-\delta).
\ee
On the other hand,  
\begin{eqnarray} 
H(X|Y) & = & H(X) - I(X;Y) \nonumber \\
& = & \log N - I(X;Y) \nonumber \\
& \ge & \log N - S(\rho_C(K))
\label{ineq1}
\end{eqnarray}
where in the last line we used Holevo's bound \cite[p.~531]{NC}.

Let $\mu_k$ be the largest eigenvalue (sup-norm) of $\rho_C(k)$.
We observe that $\mu_k$ begins at time 0 with the value 1 and evolves to the final value $\mu_K$ at the termination of the algorithm.
We have 
\be
S(\rho_C(K)) \le - \mu_K\log \mu_K - (1-\mu_K) \log [(1-\mu_K)/(N-1)] \le \binent(\mu_K) + (1-\mu_K) \log N.
\label{ineq2}
\ee
since the entropy is maximized, for a fixed $\mu_K$, 
by setting the remaining $N-1$ eigenvalues equal to $(1-\mu_K)/(N-1)$.
Combining (\ref{ineq1}) and (\ref{ineq2}),
\be
\mu_K \log N \le P_e \log N + \binent(P_e) + \binent(\mu_K) \le P_e \log N + 2
\label{supbound}
\ee
Now, let 
\be
\Delta = \sup \{|\mu_{k+1}-\mu_{k}|:\; k=0,1,\ldots,K-1\}.
\ee
This is the maximum change in the sup norm of $\rho_C(k)$ per algorithmic step.
Clearly,
\[
K \ge \frac{1-\mu_K}{\Delta}.
\]
Using the inequality (\ref{supbound}), we obtain
\be
K \ge \frac{1-P_e + 2/\log N}{\Delta}.
\ee
Thus, any upper bound on $\Delta$ yields a lower bound on $K$.
The proof will be completed by proving
\begin{Lemma} $\Delta \le 2\pi / \sqrt{N}$.
\end{Lemma}

We know that operators that do not involve oracle calls do not change the 
eigenvalues, hence the sup norm, of $\rho_C(k)$.
So, we should only be interested in bounding the perturbation of the eigenvalues of $\rho_C(k)$ as a result of an oracle call.
We confine our analysis to the oracle operator (\ref{oracleoperator}) that the  Grover algorithm uses.

For purposes of this analysis, we shall consider a continuous-time representation for the operator $O_x$ so that we may break the action of 
$O_x$ into infinitesimal time steps. 
So, we define the Hamiltonian
\be
H_x = - \pi \rx\lx
\ee
and an associated evolution operator
\[
O_x(\tau) = e^{-i\tau H_x} = I + (e^{i\pi \tau} - 1) \rx\lx.
\]
The operator $O_x$ is related to $O_x(\tau)$ by 
$O_x=O_x(1)$.

We extend the definition of conditional density to continuous time by 
\be
\rho_x(k_0+\tau) = O_x(\tau) \rho_x(k_0) O_x(\tau)^\dagger
\ee
for $0\le \tau\le 1$.
The computer state in continuous-time is defined as
\be
\rho_C(t) = \sum_x (1/N)\rho_x(t).
\label{computerdensityCT}
\ee
Let $\{\lambda_n(t), u_n(t)\}$, $n=1,\ldots,N$, be the eigenvalues and associated normalized eigenvectors of $\rho_C(t)$. 
Thus,
\be
\rho_C(t) |u_n(t) \rangle = \lambda_n(t) |u_n(t)\rangle, \;\;
\langle u_n(t)| \rho_C(t) = \lambda_n(t) \langle u_n(t)|, \;\;
\langle u_n(t)| u_m(t)\rangle = \delta_{n,m}.
\ee
Since $\rho_C(t)$ evolves continuously, so do
$\lambda_n(t)$ and $u_n(t)$ for each $n$.

Now let $(\lambda(t),u(t))$ be any one of these eigenvalue-eigenvector pairs. 
By a general result from linear algebra (see, e.g., Theorem 6.9.8 of Stoer and Bulirsch \cite[p.~389]{Stoer} and the discussion on
p.~391 of the same book),
\be
\frac{d\lambda(t)}{dt} = \lut \frac{d\rho_C(t)}{dt} \rut.
\label{derivative}
\ee
To see this, we differentiate the two sides of the identity $\lambda(t) = \lut \rho_C(t) \rut$, to obtain
\begin{eqnarray*}
\frac{d\lambda(t)}{dt} & =& \langle u'(t) | \rho_C(t) \rut + \lut \frac{d\rho_C(t)}{dt} \rut + \lut \rho_C(t)|u'(t)\rangle\\
& = & \lut \frac{d\rho_C(t)}{dt} \rut + \lambda(t) [\langle u'(t) \rut + \lut u'(t) \rangle] \\
& = &  \lut \frac{d\rho_C(t)}{dt} \rut + \lambda(t) \frac{d}{dt} \lut u(t)\rangle  \\
& = &  \lut \frac{d\rho_C(t)}{dt} \rut 
\end{eqnarray*}
where the last line follows since $\lut u(t)\rangle \equiv 1$.

Differentiating (\ref{computerdensityCT}), we obtain
\be
\frac{d \rho_C(t)}{dt} = \sum_x -(i/N) [H_x,\rho_x(t)]
\ee
where $[\cdot,\cdot]$ is the commutation operator.
Substituting this into (\ref{derivative}), we obtain 
\begin{eqnarray*}
\left|\frac{d\lambda(t)}{dt} \right|
& = & \left|\lut - \frac{i}{N} \sum_x [H_x,\rho_x(t)]\, \rut \right|\\
& \le &  \frac{2}{N} \left| \sum_x  \lut  H_x\rho_x(t) \rut \right|\\
& \stackrel{(a)}{\le} & \frac{2}{N}\sqrt{\sum_x \langle u(t) |H_x^2 | u(t) \rangle  } \;\sqrt{\sum_x  \langle u(t) |\rho_x^2(t) | u(t) \rangle  } \\
& \stackrel{(b)}{=} & \frac{2}{N} \sqrt{\sum_x \pi^2 \left| \langle u(t) |x \rangle \right|^2 }\;\sqrt{N\langle u(t) |\rho_C(t) | u(t) \rangle } \\
& = & \frac{2\pi}{\sqrt{N}} \;1 \cdot \sqrt{\lambda(t)} \\
& \le & \frac{2\pi}{\sqrt{N}}
\end{eqnarray*}
where $(a)$ is the Cauchy-Schwarz inequality, $(b)$ is due to (i) $\rho_x^2(t) = \rho_x(t)$ as it is a pure state, and (ii) the definition (\ref{computerdensityCT}).
Thus, 
\be
|\lambda(k_0+1) - \lambda(k_0)| = \left| \int_{k_0}^{k_0+1} \frac{d\lambda(t)}{dt} dt \right| \le 2\pi/\sqrt{N}.
\ee
Since this bound is true for any eigenvalue, the change in the sup norm of 
$\rho_C(t)$ is also bounded by $2\pi /\sqrt{N}$.

\subsection*{Discussion}
The bound (\ref{Prop1}) captures the $\sqrt{N}$ complexity of Grover's search algorithm.
As mentioned in the Introduction, lower-bounds on Grover's algorithm have been known before; and, in fact, the present bound
is not as tight as some of these earlier ones.
The significance of the present bound is that it is largely based on information-theoretic concepts.
Also worth noting is that the probability of error $P_e$ appears explicitly in (\ref{Prop1}), unlike other bounds known to us.

\end{document}